\documentclass[prl,superscriptaddress,twocolumn,floatfix,showpacs,preprintnumbers,amsmath,amssymb,amsfonts,lengthcheck]{revtex4-2}

\usepackage[dvips]{graphics}
\usepackage{newtxmath}
\usepackage[colorlinks,pdfusetitle,urlcolor=blue,citecolor=blue,linkcolor=blue,bookmarksnumbered,plainpages=false]{hyperref}

\usepackage{braket}

\usepackage[english]{babel}
\usepackage{mathtools}

\newcommand{\mi}{\textrm{i}} 
\newcommand{\me}{\mathrm{e}}

\begin{document}

\title{Quantized fields for optimal control in the strong coupling regime}

\author{Frieder Lindel}
\affiliation{\freiburg}
\author{Edoardo G.\ Carnio}
\affiliation{\freiburg}
\affiliation{\eucor}
\author{Stefan Yoshi Buhmann}
\affiliation{\kassel}
\author{Andreas Buchleitner}
\affiliation{\freiburg}
\affiliation{\eucor}

\newcommand{\freiburg}{Physikalisches Institut, Albert-Ludwigs-Universit\"{a}t Freiburg, Hermann-Herder-Stra{\ss}e 3, 79104, Freiburg, Germany}
\newcommand{\eucor}{EUCOR Centre for Quantum Science and Quantum Computing, Albert-Ludwigs-Universit\"{a}t Freiburg, Hermann-Herder-Stra{\ss}e 3, 79104, Freiburg, Germany}
\newcommand{\kassel}{Institut f\"{u}r Physik, Universit\"{a}t Kassel, Heinrich-Plett-Stra{\ss}e 40, 34132 Kassel, Germany}

\date{\today}

\begin{abstract}
We tailor the quantum statistics of a bosonic field to deterministically drive a quantum system into a target state. Experimentally accessible states of the field achieve good control of multi-level or -qubit systems, notably also at coupling strengths beyond the rotating-wave approximation. This extends optimal control theory to the realm of fully quantized, strongly coupled control and target degrees of freedom.
\end{abstract}

\maketitle

The realization of quantum-based technologies (sensors, computers, simulation platforms) requires the isolation and manipulation, i.e., the control of their quantum-mechanical constituents. Control means to use an external degree of freedom (a field) to impart some desired dynamics, i.e., to drive the target constituent from a given initial to a desired final state. If the driving is coherent, we talk of \emph{coherent control} \cite{shapiro_quantum_2011}. The search for the state of a control field that maximizes an objective function -- a desired property which constrains the state transformation -- defines the field of \emph{optimal} control \cite{assion_control_1998, glaser_training_2015}, which 
finds applications in diverse areas such as molecular physics \cite{Brumer1986,Tannor1986,Shi1988}, semiconductor quantum structures \cite{kosionis2007optimal}, Bose-Einstein condensates \cite{hohenester2007optimal}, open quantum systems \cite{Schmidt2011,Sauer2013}, or multipartite entanglement \cite{Platzer2010} generation, e.g., between NV centres \cite{Dolde2014}.

The natural question arises whether these protocols can be improved by exploiting the quantum nature of the control field \cite{carreno_excitation_2016,carreno_excitation_2016-1,mukamel_roadmap_2020}. First steps have been taken in the optimization of multi-photon excitation processes which are determined by higher-order coherences \cite{lambropoulos_field-correlation_1968,agarwal_field-correlation_1970}. For single-mode fields, nonclassical squeezed light can drive these processes more efficiently than coherent light of the same intensity \cite{gea-banacloche_two-photon_1989,georgiades_nonclassical_1995,spasibko_multiphoton_2017, li_squeezed_2020}. These effects have been exploited, e.g., to improve protocols for quantum sensing \cite{lopez_carreno_exciting_2015,munoz_quantum_2021} or spectroscopy \cite{raymer_entangled_2013,dorfman_nonlinear_2016,basset_perspectives_2019,szoke_entangled_2020,raymer_entangled_2021}. Further enhancement may be achieved by making use of quantum correlations in the control field, e.g., in multimode quantum light. For low-intensity light sources, two-photon states entangled in frequency and in time \cite{Hong1987} have been shown to bear significant advantages over classically shaped pulses in the control of two-photon absorption \cite{lee_entangled_2006, upton_optically_2013,schlawin_theory_2017,carnio_optimization_2021,carnio_how_2021}. In general, the weak interactions that define the perturbative regime limit the means to exert control to the shaping of $ n $-point correlation functions of the field \cite{landes_experimental_2021}.

The more intricate dynamics of \emph{strongly} coupled control and target offer a broader range of possible optimization paths. Recent technological progress, 
achieving strong coupling in a multitude of platforms \cite{forn-diaz_ultrastrong_2019,kockum_ultrastrong_2019}, has led to the nascent fields of polaritonic chemistry \cite{galego_cavity-induced_2015,herrera_cavity_controlled_2016,flick_strong_2018} and quantum materials \cite{schlawin_cavity_2022}. Here, the formation of hybridized light-matter 
states (polaritons) is used, e.g., to alter the chemical properties of molecules \cite{hutchison_modifying_2012,galego_suppressing_2016}, to modify charge transport \cite{orgiu_conductivity_2015}, and to enhance collective effects \cite{schlawin_cavity_mediated_2019,li_manipulating_2020,appugliese_breakdown_2022} in solids. While, so far, the molecular or solid targets are often resonantly coupled to the vacuum of the field, active design of the latter's state has been discussed only rarely \cite{Wu2019,csehi_quantum_2019,castro_optimal_2019}.

\begin{figure}[b]
\includegraphics[width=0.8\columnwidth]{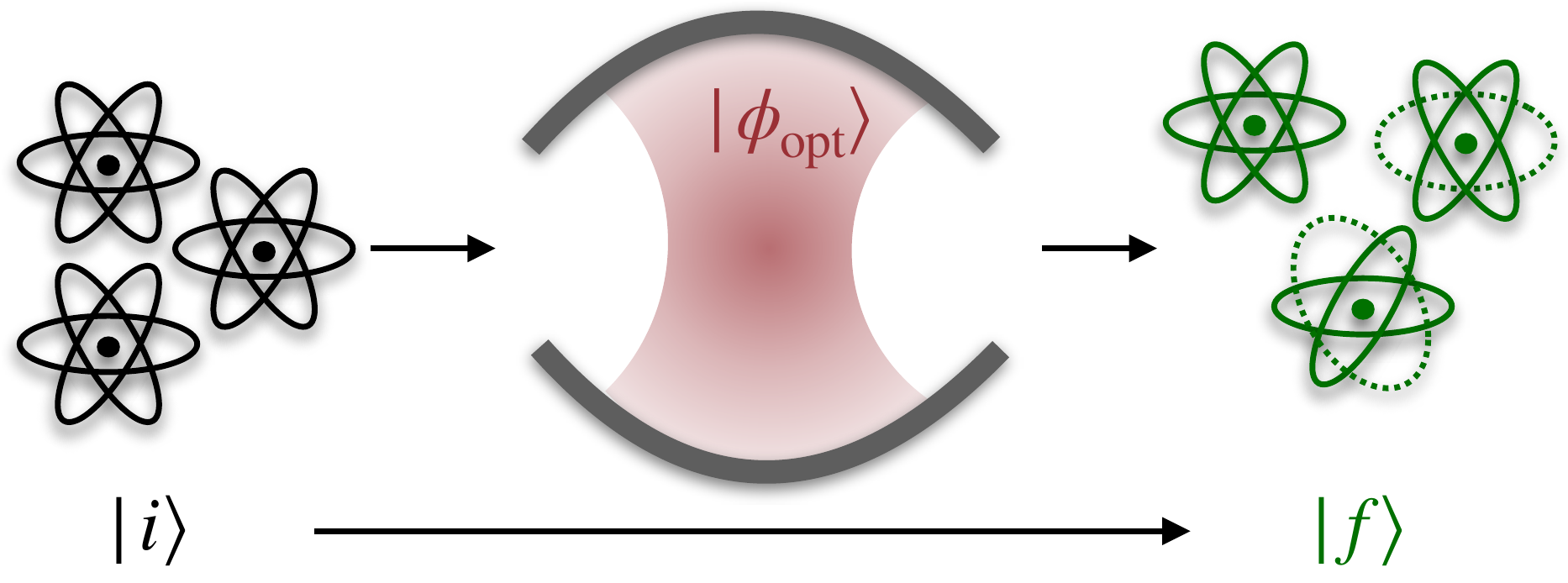}
\caption{\textit{Optimal control problem under consideration.} Given a quantum system (black atoms) initially prepared in state $\ket{i}$ and interacting with a bosonic field (red shaded cavity field), which is the optimal initial state of the field $\ket{\phi_\text{opt}}$ such that, after an interaction time $T$, the sample is in a predefined target state $\ket{f}$ (green atoms)?}
\label{fig:Scheme}
\end{figure}

In our present contribution, we employ an optimal quantum control theory to infer the initial state of a bosonic field mode, with tailored quantum statistics, to optimally drive a multi-level quantum system from its initial into a desired target state, see Fig.~\ref{fig:Scheme}.
Our second-quantized control scheme is valid across all coupling strengths (no rotating-wave approximation), including regimes which are hard to tackle analytically, and which so far have been deemed less suitable for specific control problems \cite{castro_optimal_2019}.
In contrast, we show that, even in these latter regimes, we can find experimentally realizable optimal control states. Our findings apply to various scenarios such as molecules in optical \cite{schwartz_reversible_2011} or plasmonic cavities \cite{chikkaraddy_single-molecule_2016}, superconducting circuits \cite{blais_circuit_2021}, electron transfer reactions \cite{may_charge_2011}, or trapped ion systems \cite{HAFFNER2008}. They thus hold the promise to become a relevant tool in a broad range of contexts. 

Given the \emph{quantized} control field and the target system initially in a product state, which is the optimal initial state 
$ \ket{\phi_{\rm opt}} $ of the field, such that the 
probability to find the target in a desired final state after a fixed interaction (or control) time $T$ is maximal? The Hamiltonian $ \hat{H} $ of the joint system determines, via the Schr\"odinger equation, the time evolution operator $\hat{U}(t, t_0 = 0)$. Initial $ \ket{i} $ and final $ \ket{f} $ states of the target fix the transition operator $ \hat{T}_{fi}(t) = \braket{f | \hat{U}(t,0) | i}$ acting upon the control field's Hilbert space $ \mathcal{H}_c $. 
With the additional constraint that the control field's excitation be bound from above by $N_{\rm max}$ (to be able to optimize over a finite-dimensional
space, see below), implemented by the projector $ \hat{\Pi} $ on the corresponding subspace of $ \mathcal{H}_c $, the 
target state population reads $ p_f(t) = \braket{\phi_i | \hat{T}_{fi}^\dagger(t) \hat{\Pi} \hat{T}_{fi}(t) | \phi_i}$, with $ \ket{\phi_i} $ the initial state of the control field.  \ To determine the (normalized) state $ \ket{\phi_{\rm opt}} $ that maximizes $ p_f(t) $ at time $ T $, we introduce the Lagrange multiplier $ \lambda $ and define the functional $ J[\ket{\phi_i}, \bra{\phi_i}, \lambda] = p_f(T) -\lambda \left(\braket{\phi_i|\phi_i}-1\right) $. Maximization \cite{schlawin_theory_2017,carnio_optimization_2021,carnio_how_2021} of $ J $ yields the eigenvalue problem
\begin{align} \label{eq:EigenvalueProblem}
\hat{\mathcal{M}} \ket{\phi_i} = \lambda \ket{\phi_i},
\end{align}
with $ \hat{\mathcal{M} }= \hat{T}_{fi}^\dagger(T) \hat{\Pi} \hat{T}_{fi}(T) $ fully defining the control problem. The largest eigenvalue $ \lambda_\text{opt} $ of \eqref{eq:EigenvalueProblem} equates the \emph{fidelity} $ \mathcal{F} $ of the control, 
i.e., the maximally achievable probability to find the target in the desired state, upon driving it with a control field prepared in the eigenstate $ \ket{\phi_\text{opt}}$ pertaining to $ \lambda_\text{opt} $.
\begin{figure}
\includegraphics[width=1\columnwidth]{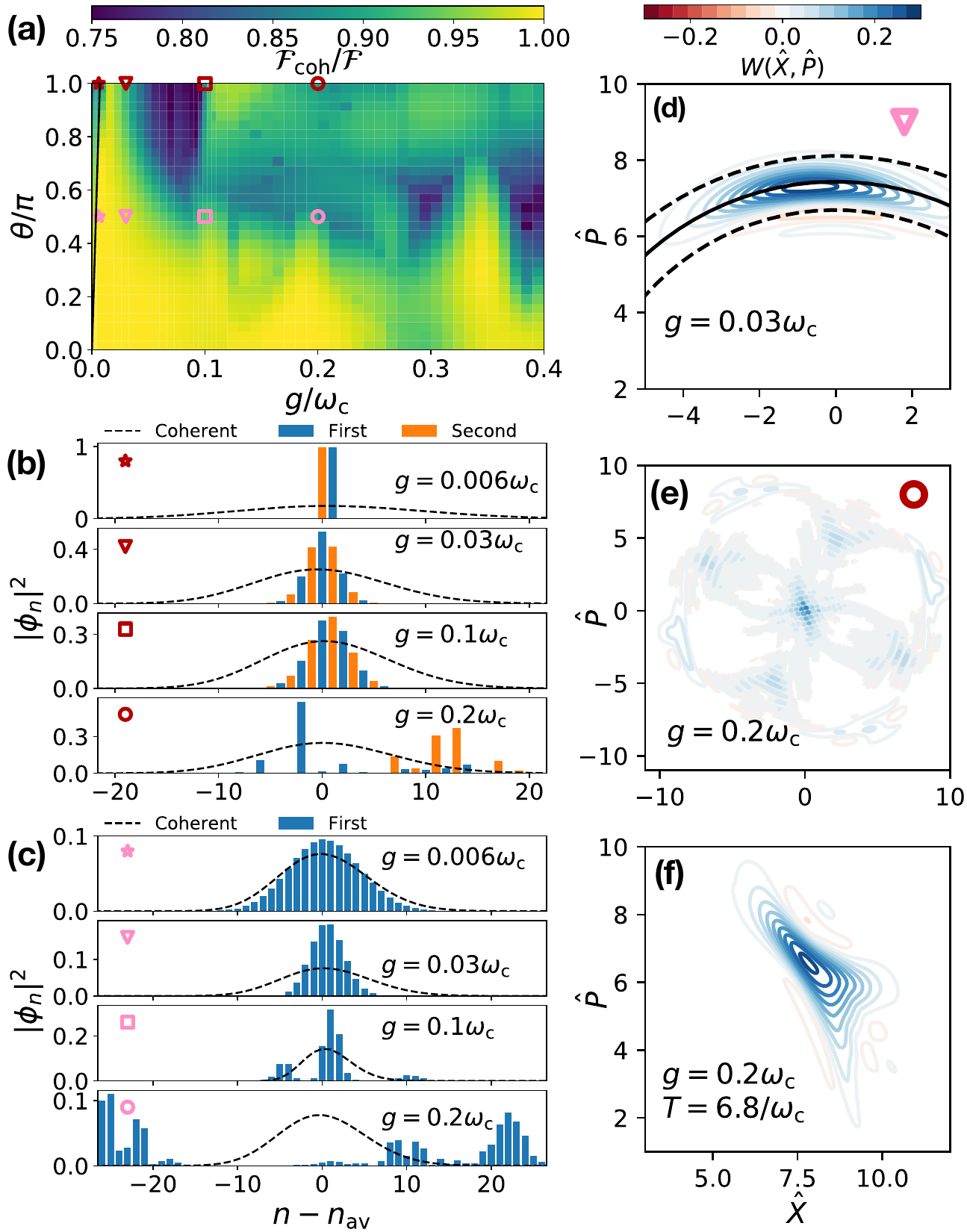}
\caption{\textit{Optimal control of a single resonant two-level system via a bosonic mode.}  (a) Ratio between the fidelities achieved using the optimal coherent state ($ \mathcal{F}_\text{coh} $), and the solution of the optimal control problem ($ \mathcal{F}$), as a function of the coupling strength $g$ and the target state's Bloch angle $\theta$ ($\varphi=0$), for an interaction time $T = 25/\omega_c $. The solid black line indicates $2\sqrt{N_\mathrm{max}} g T =\theta$.  (b, c) Number statistics $\phi_n = \braket{n | \phi_\text{opt}} $ of the two optimal states [associated with the two largest eigenvalues of \eqref{eq:EigenvalueProblem}], for different coupling strengths (top to bottom), $T = 25/\omega_c $, 
and final states with $\theta = \pi, \pi/2$ in (b) and (c), respectively. The black dashed lines correspond to  Poissonian distributions (multiplied by magnification factors to ease the comparison of the distribution's width) with average photon numbers $n_\text{av}$ equal to those of the optimal states. (d, e, f) Wigner functions of the optimal states for $T = 25/\omega_c $, $\theta = \pi/2$ and $g=0.03 \omega_c$ (d), $T = 25/\omega_c $, $\theta = \pi$ and $g=0.2 \omega_c$ (e), and $ T = 6.8/\omega_c $, $\theta = \pi/2$, and $g=0.2 \omega_c$ (f). In (d) the black solid and dashed lines correspond to circles 
around the origin with radius $n_\text{av}$ and $n_\text{av} \pm \sqrt{n_\text{av}}$, respectively, highlighting the sub-Poissonian nature of the state in radial direction.}
\label{fig:TwoLevelSystem}
\end{figure}

Hereafter, we apply the optimal control protocol above to different scenarios, all described by the Hamiltonian (in units of $ \hbar $)
\begin{multline} \label{eq:Hamiltonian}
\hat{H} =  \omega_c \hat{a}^\dagger \hat{a} 
+\sum_{j = 1}^{N_A}\sum_{k = 1}^{N^{(j)}_L} \omega_k^{(j)} \hat{\sigma}^{(j)}_{kk}  +\sum_{j=1}^{N_A} \sum_{\substack{ k \neq l=1}}^{N_L^{(j)}} g_{kl}^{(j)}  \hat{\sigma}^{(j)}_{kl} \hat{X}\, ,
\end{multline} 
which features linear coupling of strength $ g_{kl}^{(j)} $ between a bosonic mode of frequency $\omega_c$, with associated annihilation and creation 
operators $ \hat{a},\hat{a}^\dagger $, and the bound states at energies $\omega^{(j)}_k$, $k = 1, \ldots,  N_L^{(j)}$, of $j=1,\ldots,N_A$ atoms, mediated by the atomic 
transition operators $\hat{\sigma}^{(j)}_{kl}$, together with the 
control field amplitude $\hat{X} = (\hat{a} + \hat{a}^{\dagger})/2$. In particular, (\ref{eq:Hamiltonian}) includes ``counter-rotating'' terms (\textit{\`a la} $ \hat{\sigma} \hat{a}$, $\hat{\sigma}^\dagger \hat{a}^\dagger $) that become relevant for large couplings and detunings $ \vert \omega_k^{(j)} - \omega_l^{(j)} \vert - \omega_c  $.

We first apply our control scheme to a single two-level atom with ground state $\ket{1} = \ket{G}$ ($ \omega_G = 0 $) and excited state $\ket{2} = \ket{E}$, such that  $N_A = 1$, $N_L = 2$, and Eq.~\eqref{eq:Hamiltonian} reduces to the Rabi Hamiltonian \mbox{$\hat{H} = \omega_c \hat{a}^\dagger \hat{a}  +  \omega_E \hat{\sigma}_{EE} +g \hat{\sigma}_x\left( \hat{a} +\hat{a}^\dagger\right) $} \cite{rabi_process_1936}, with $\hat{\sigma}_x =  \hat{\sigma}_{GE} + \hat{\sigma}_{EG}$. We analyze the resonant case, $\omega_c = \omega_E$, with the goal to drive the system, within time $ T $, from the initial state $ \ket{G} $ to an arbitrary target state $\ket{f} = \cos( \theta/2) \ket{G} + \me^{\mi \varphi} \sin(\theta/2) \ket{E}$, parametrized by angles $\theta$ and $\varphi$ on the Bloch sphere. The optimal state of the control field  is obtained by diagonalization of $\hat{\mathcal{M}}$ in Eq.~\eqref{eq:EigenvalueProblem}, with $\hat{\Pi} = \sum_{n=0}^{N_\mathrm{max}} \ket{n}\bra{n}$ and $\ket{n}$ the field's Fock states.

To illustrate the added value of a quantized control field, Fig.\ \ref{fig:TwoLevelSystem}(a) contrasts the target state fidelity $ \mathcal{F} $ achieved by $ \ket{\phi_{\rm opt}} $ against the best fidelity $\mathcal{F}_\text{coh} = \braket{\alpha | \hat{\mathcal{M}} | \alpha}$ obtained by a (classical) coherent state $ \ket{\alpha} $ with average 
photon number $ \vert \alpha \vert^2 \leq N_\text{max} $, for $T = 25/\omega_c$, $N_\text{max} = 80$, and $\varphi = 0$ (other choices do not affect the observations that follow).
For $2gT\sqrt{N_\mathrm{max}} > \theta$ [right of the solid black line on the very left of Fig.~\ref{fig:TwoLevelSystem}(a)], i.e., when the coupling is strong enough for the field to invert the population of the target within the control time, we can always find an optimal solution that yields the target state with fidelity $\mathcal{F} > 0.98$ (absolute values of $\mathcal{F}$ 
not shown in Fig.~\ref{fig:TwoLevelSystem}).
If $ g \ll \omega_c $ and $ g T < 1 $, we are in the realm of classical coherent control, and therefore classical-like coherent states perform as well as the optimal solutions for all $ \theta $, i.e., $\mathcal{F}_\text{coh} /\mathcal{F}\approx 1$ [see the yellow area by the black line in Fig.~\ref{fig:TwoLevelSystem}(a), extending over the entire interval of Bloch angles from $\theta = 0$ to $\theta=\pi$].
If $ gT > 1 $, however, the Rabi oscillations induced by coherent states collapse \cite{Mandel1995} before the target state is fully populated, and coherent states typically offer worse control than the optimal solutions of Eq.~\eqref{eq:EigenvalueProblem}, as indicated by green and blue areas in Fig.~\ref{fig:TwoLevelSystem}(a).

To understand the advantage brought about by the optimal states $ \ket{\phi_\text{opt}} $ for $\ket{f} = \ket{E}$, we plot their coefficients $ \phi_n = \braket{n | \phi_\text{opt}} $ in the Fock basis, in Fig.\ \ref{fig:TwoLevelSystem}(b).
For small coupling strengths ($ g \ll \omega_c $) we are in the realm of the Jaynes-Cummings Hamiltonian (recall $ \omega_E = \omega_c $) \cite{Mandel1995}, a non-perturbative regime where the rotating-wave approximation (RWA) applies and the Hamiltonian \eqref{eq:Hamiltonian} decomposes into blocks identified by the total number of excitations shared between control and target degrees of freedom. Within each of these blocks, the system oscillates \cite{Mandel1995} between states $ \ket{G}\ket{n} $ and $ \ket{E}\ket{n-1} $, with Rabi frequency $ \Omega_n = 2\sqrt{n} g$. A Fock state with $ n $ excitations, such that $ 2\sqrt{n} g T =  \pi (2 k + 1) , k \in \mathbb{N}$, is therefore sufficient to drive the target system from $ \ket{G} $ to $ \ket{E} $ within the time interval $ T $. This is shown in the first row of Fig.~\ref{fig:TwoLevelSystem}(b).

This simple intuition fails if we want to drive the transition $ \ket{G} \rightarrow (\ket{G} + \ket{E})/\sqrt{2}$: it is not sufficient to pick a Fock state $ \ket{n} $ satisfying $  2\sqrt{n} g T = \pi k + \pi/2 $, since, at the control time, control and target will be entangled, and the atomic subsystem in the (manifestly undesired) maximally mixed state $ (\ket{G}\bra{G} + \ket{E}\bra{E})/2 $. Indeed, a fundamental requirement of the control protocol is that target and control be \emph{uncorrelated} at the control time \cite{wellens_quantum_2000}, and therefore $ \ket{\phi_\text{opt}} $ must contain the necessary components to cancel any entanglement built up during the interaction. This is achieved by a state with the number distribution shown in the first row of Fig.\ \ref{fig:TwoLevelSystem}(c) -- which is sub-Poissonian, even for small values of $gT$, as evidenced by the comparison with the Poissonian profile represented by the black dashed line. The non-classical nature of the optimal state is particularly visible in the second row of Fig.~\ref{fig:TwoLevelSystem}(c), and from the corresponding Wigner function in Fig.~\ref{fig:TwoLevelSystem}(d): these two representations indicate that the optimal states in this regime are number-squeezed states \cite{kitagawa_number-phase_1986}.

\begin{figure}
\includegraphics[width=1\linewidth]{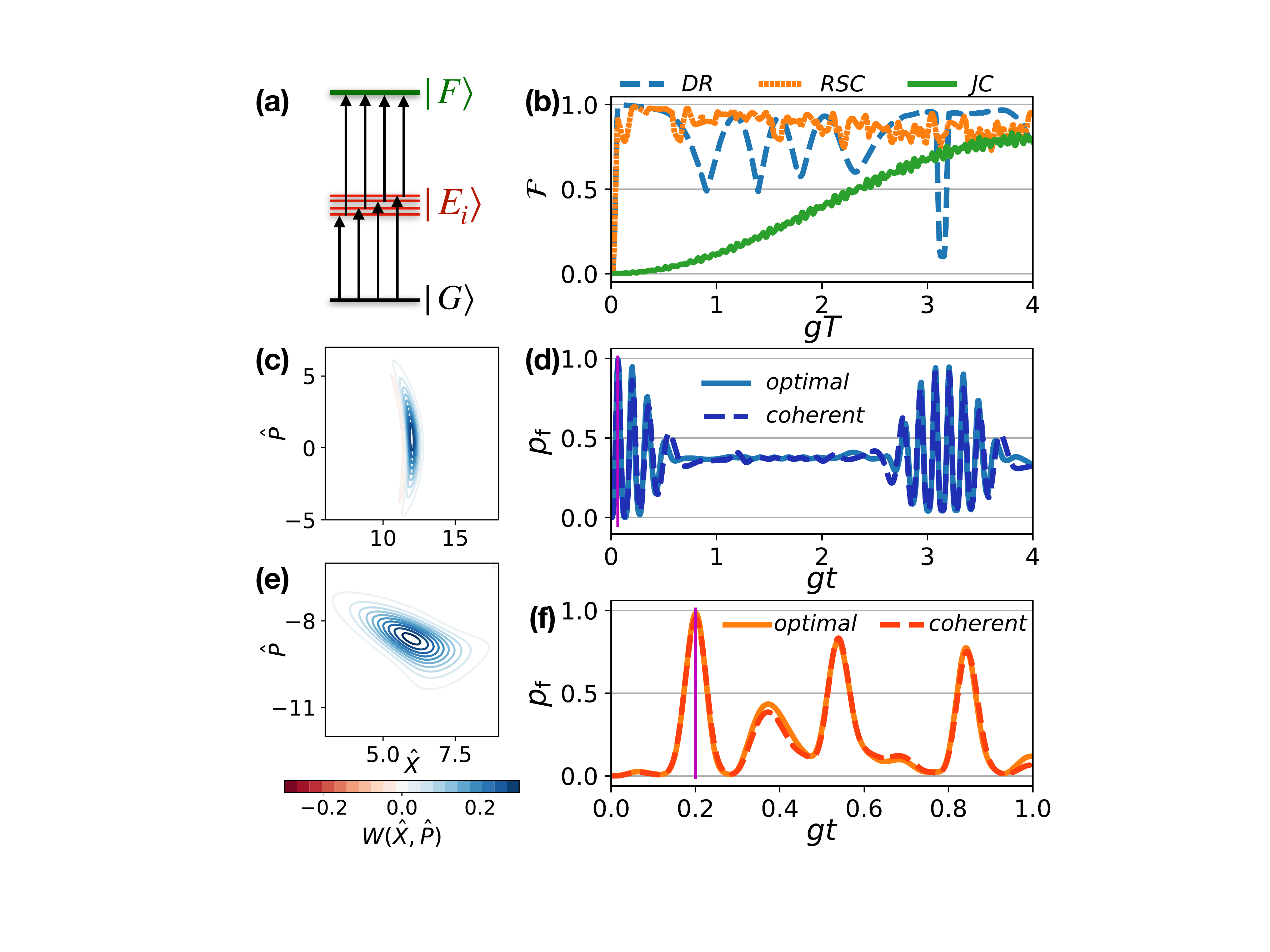} 
\caption{\textit{Control of a multi-level system.} (a) Target level scheme under consideration, with arrows indicating the allowed transitions with $g_{kl}^{(1)} \neq 0$ in \eqref{eq:Hamiltonian}. (b) Fidelities $\mathcal{F}$ achieved for three different parameter ranges: Jaynes-Cummings (JC; $g=0.1 \omega_c$, $\omega_{E_1} =0.86 \omega_c $, $\omega_{E_2} = 0.92 \omega_c$, $\omega_{E_3} =1.1 \omega_c$, $\omega_{E_4} =1.2 \omega_c$, $\omega_{F} = 2 \omega_c $), resonant strong coupling (RSC; $g=0.1 \omega_c$, spectrum as for JC), and diabatic (DR; $g=0.5\omega_c$, $\omega_{E_1} = 0.32\omega_c$, $\omega_{E_2} = 0.35\omega_c$, $\omega_{E_3} =0.4 \omega_c$, $\omega_{E_4} = 0.42 \omega_c$, $\omega_{F} = 0.5\omega_c$). Wigner function (c) of the optimal state for $T = 0.02 \times 2 \pi/\omega_c$, and (d) population of the target state $\ket{F}$, as a function of time, respectively, in the DR. In (d), the control field is prepared either in the optimal, or in the coherent state $\ket{\alpha}$ with $\alpha =  \bra{\phi_\mathrm{opt}} \hat{X} + \mi \hat{P} \ket{\phi_\mathrm{opt}} $, and the vertical purple line indicates the control time $T$. (e, f): Same as in (c) and (d), respectively, but for the resonant strong coupling case (RSC), and $T = 0.01 \times 2 \pi/\omega_c$. $N_\mathrm{max}=80$, in all cases.}
\label{fig:Multilevel}
\end{figure}

With increasing coupling strength $g$, more and more manifolds, with even or odd excitation numbers, are coupled by the counter-rotating terms, and become available to the optimization algorithm to build the optimal solutions, as visible in the lower parts of Fig.~\ref{fig:TwoLevelSystem}(b,c). While the structure of these optimal states in phase space is in general complex [Fig.~\ref{fig:TwoLevelSystem}(e)], we find combinations of $g$ and $T$ for which optimal states \footnote{For problems with conserved parity, the eigenstates of $ \hat{\mathcal{M}} $ have themselves well-defined even or odd parity [compare blue and orange bars in Fig.\ \ref{fig:TwoLevelSystem}(b)]. For the cases presented in Fig.\ \ref{fig:TwoLevelSystem}(d,f) and \ref{fig:Multilevel}(c,e), the first two optimal solutions are almost degenerate in $ \mathcal{F} $, and opposite in parity. We can therefore superimpose these states to find localized solutions in phase space.} are well localized in phase space [Fig.\ \ref{fig:TwoLevelSystem}(f)] and may be approximated by experimentally accessible states such as coherent and squeezed states. In the following we show that such localized optimal solutions are also found for the control of more complicated target systems. 

Let us consider the multilevel system illustrated in Fig.~\ref{fig:Multilevel}(a). 
Its spectrum resembles that of a molecule with an energetically isolated target state $\ket{F}$ that can only be reached from the ground state $\ket{G}$ via a manifold of four intermediate levels $\ket{E_i}$, $i = 1,\dots, 4$. We assume that all allowed transitions have the same coupling strength $g$, which does not limit the generality of our conclusions below. We analyse three different parameter ranges: 
\begin{itemize}
\item[(i)] the Jaynes-Cummings (JC) regime, with near-resonant target and cavity transitions, and \mbox{$g = 0.001 \omega_c$}, such that the RWA applies;
\item[(ii)] the resonant strong coupling (RSC) regime \footnote{This regime, in which the coupling constant $g$ is comparable to the bare frequencies, is also referred to as `ultra-strong coupling' \cite{forn-diaz_ultrastrong_2019,kockum_ultrastrong_2019}, to distinguish it from the `strong coupling' regime \cite{thompson1992observation} where the coherent dynamics is faster than dissipative processes.}, with near-resonant transitions as in the JC regime, but with $g=0.1 \omega_c$, so that the RWA does not apply \cite{shirley65};
\item[(iii)] the diabatic regime (DR), where the target's transition frequencies are smaller than the cavity frequency ($\omega_{E_{j}} , \omega_F - \omega_{E_{j}} < \omega_c$) \footnote{In the context of energy transfer in molecular physics, where fast vibrational modes (which play the role of the control field) are coupled to slower electronic states, this regime is also called `non-adiabatic' \cite{may_charge_2011}. Note, however, that other communities call this regime `adiabatic' \cite{irish_dynamics_2005}.}. Here $g=0.5\omega_c$.
\end{itemize}

In Fig.~\ref{fig:Multilevel}(b) we show the achievable target fidelities $\mathcal{F}$ as a function of the control time $T$: in the JC regime, several Rabi cycles ($2gT \sqrt{N_\mathrm{max}}\gg \pi$) are necessary to reach $\mathcal{F} > 0.8$, since all intermediate levels are detuned from the cavity frequency (the typical detuning $ \Delta \gg g \sqrt{N_\text{max}} $), leading to an incomplete population transfer from the ground to the excited states \cite{Mandel1995}.

In the diabatic and RSC regimes, instead, $ \Delta \ll g \sqrt{N_\text{max}} $ due to the stronger coupling; therefore, the detunings do not affect the time evolution of the target on timescales determined by $2gT \sqrt{N_\mathrm{max}} \approx \pi$.	Interestingly, it is on these short timescales that control becomes possible: we observe $\mathcal{F} \approx 1$ already for $gT \approx 0.1, 0.2$, respectively, compare Fig.\ \ref{fig:TwoLevelSystem} (b). This makes the diabatic and RSC regimes well-suited for the deterministic driving of non-trivial level structures. What is more, this fast control is achieved by optimal states \cite{Note1} of the field which are well localised in phase space and reminiscent of squeezed states, compare Figs.~\ref{fig:Multilevel}(c) and (e) for the optimal states obtained for $T=0.01\times 2\pi/\omega_c$ (DR) and $T=0.2/\omega_c$ (RSC), respectively. This is remarkable, since, in general, optimal states exhibit complicated statistics in the strong coupling regimes, compare Fig.~\ref{fig:TwoLevelSystem}(e). The optimal states in Figs.~\ref{fig:Multilevel}(c) and (e) can be approximated by coherent states $\ket{\alpha}$, with $\alpha = \bra{\phi_\mathrm{opt}} \hat{X} + \mi \hat{P} \ket{\phi_\mathrm{opt}}  $, $\hat{P} = (\hat{a} - \hat{a}^{\dagger})/2\mi$. In Figs.~\ref{fig:Multilevel}(d) and (f) we show the population of the target state, as a function of time $t$, when the field is initially prepared in $\ket{\phi_\mathrm{opt}}$ or in $\ket{\alpha}$.
We find that, in both regimes, also the coherent states drive the system with high fidelity (DR: $\mathcal{F}_\text{coh} = 0.99$; RSC: $\mathcal{F}_\text{coh} = 0.98$) into target, at time $T$. 

These results demonstrate that control with experimentally accessible field states can be achieved beyond the rotating-wave approximation, even for a more complicated target level structure. The dynamics in the DR [Fig.~\ref{fig:Multilevel}(d)] display the well-known collapses and revivals of the populations in the initial state at $t =2\pi/\omega_c$ [$gt =\pi$ in Fig.~\ref{fig:Multilevel}(d)]\cite{casanova_deep_2010}. In the RSC regime, the distinct structure of collapses and revivals signifies the departure from the DR. 
%
\begin{figure}
\includegraphics[width=1\linewidth]{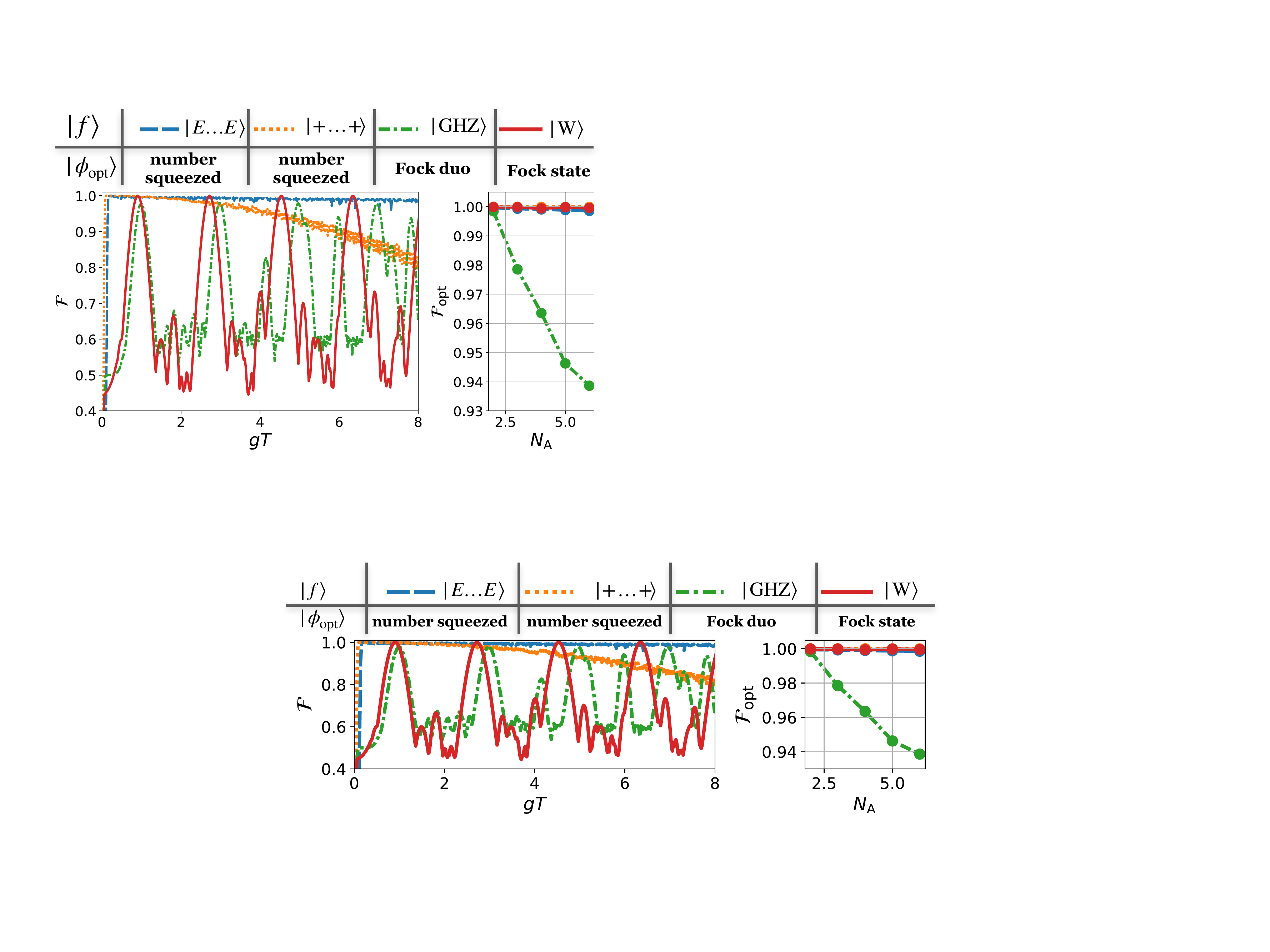} 
\caption{\textit{Control of multiple qubits.} Table: quantum statistics of the optimal state $\ket{\phi_\mathrm{opt}}$ driving multiple qubits from their ground state to different final states $\ket{f}$ (see main text). Left: target fidelity $\mathcal{F}$ vs.\ control time $T$, for $ g = 0.01 \omega_c $ and $N_\mathrm{A} = 3$ qubits. Right: largest target fidelity $ \mathcal{F}_\text{opt} $ achievable within $0 \leq gT \leq 8 $, vs. register length $N_A$ (same colour code as on the left; target fidelities of all but the GHZ state essentially coincide).}
\label{fig:MultipleQubits}
\end{figure}

As a last example, let us consider $N_A$ two-level systems ($N_L = 2$) or qubits, each coupled to the same, single field mode, with coupling constant $g = 0.01 \omega_c$. We want to steer the atoms into (a) the GHZ state $\ket{\mathrm{GHZ}} = (\ket{G\dots G} + \ket{E \dots E})/\sqrt{2}$, (b) the W state $\ket{\mathrm{W}} = \mathrm{sym}(\ket{E G G \dots })/\sqrt{N_A}$ ($\mathrm{sym}(x)$ symmetrizes the state $x$), (c) a state with all qubits excited, $\ket{E\dots E}$, and (d) with all in $ \ket{+} = (\ket{E} + \ket{G})/\sqrt{2} $, i.e., $\ket{+ \ldots +}$. The fidelity $\mathcal{F}$ is found as a function of the control time, in Fig.~\ref{fig:MultipleQubits}. For GHZ and W target states, good control is obtained only at periodic, discrete times, whereas $\ket{+ \ldots +}$ can be reached only for small values of $gT$, e.g.,  $gT<2$, to find $\mathcal{F}>0.98$ for $N_A=3$. The number statistics of the optimal states are summarized in the table in Fig.~\ref{fig:MultipleQubits}. The optimal states to populate the W state are Fock states, as discussed before \cite{castro_optimal_2019}, whereas $\ket{E\dots E}$ and $\ket{+ \ldots +}$ can be reached by number squeezed states, similarly to Fig.~\ref{fig:TwoLevelSystem}(d). It has been shown previously \cite{cirac_preparation_1994} and is here confirmed that GHZ states are prepared with Fock duos $(\ket{0} + \ket{N_A})/\sqrt{2}$.  

In conclusion, we solved the optimal control problem \eqref{eq:EigenvalueProblem} for multilevel systems driven by a quantized single mode, and found experimentally accessible optimal control states, even when the dynamics of control and target system turn intricate due to their strong coupling. The simple structure of these optimal states suggests that analytical solutions may possibly be deduced, in suitably chosen limiting cases. Our results generalize optimal control theory for strongly coupled quantum systems, and pave the way for studies of the control of more complicated molecular or condensed-matter targets.

\begin{acknowledgments}
	We acknowledge support from the Georg H.~Endress foundation (E.G.C.), and by the Studienstiftung des deutschen Volkes (F.L.).
\end{acknowledgments}


%

\end{document}